\documentclass[apps,prl,twocolumn,armsfonts,amssymb,armsmath,showpacs,floatfix,nofootinbib,groupedaddress,superscriptaddress,citesort]{revtex4}
\usepackage{mathrsfs}
\usepackage{amsfonts}
\usepackage{amstext}
\usepackage{amsmath}
\usepackage{amssymb}
\usepackage{float}
\usepackage[usenames]{xcolor}
\usepackage[dvips]{graphicx}

\def\ra{\rangle}
\def\la{\langle}

\def\no{\nonumber}
\def\bea{\begin{eqnarray}}
\def\eea{\end{eqnarray}}
\def\be{\begin{equation}}
\def\ee{\end{equation}}

\begin{document}
\title{Notes on teleportation in an expanding space}

\author{Jun Feng}

\affiliation{Beijing National Laboratory for Condensed Matter Physics,
Institute of Physics, Chinese Academy of Sciences, Beijing 100190, P. R. China}
\author{Wen-Li Yang}
\affiliation{Institute of Modern Physics, Northwest University, Xian 710069, P. R. China}
\author{Yao-Zhong Zhang}
\affiliation{School of Mathematics and Physics, The University of Queensland, Brisbane, Qld 4072, Australia}
\author{Heng Fan}
\email{hfan@iphy.ac.cn}
\affiliation{Beijing National Laboratory for Condensed Matter Physics,
Institute of Physics, Chinese Academy of Sciences, Beijing 100190, P. R. China}

\begin{abstract}

We investigate the quantum teleportation between a conformal detector Alice and an inertial detector Bob in de Sitter space in two schemes, (i) one uses free scalar modes and (ii) one utilizes cavity to store qubit. We show that the fidelity of the teleportation is degraded for Bob in both cases. While the fidelity-loss is due to the Gibbons-Hawking effect associated with his cosmological horizon in the scheme (i), the entanglement decreases in the scheme (ii) because the ability to entangle the cavities is reduced by the spacetime curvature. With a cutoff at Planck-scale, comparing with the standard Bunch-Davies choice, we also show that the possible Planckian physics cause extra modifications to the fidelity of the teleportation protocol in both schemes. \\

\end{abstract}
\maketitle

One of the main challenges in modern physics is to find a complete theory of quantum gravity which merges quantum mechanics and general relativity into a unified framework. Evidences from some candidate theories (e.g. string theory) have shown \cite{qg} that typical quantum gravitational phenomena should be a unitary process. This however conflicts with the semiclassical analysis that predict the information loss during the process. To resolve this paradox, it has become increasingly clear \cite{thooft} that non-locality, the basic feature of quantum information theory, should be employed. This indeed allows one to understand some quantum gravitational effects in a quantum-information framework. Recently, a new fast growing field called Relativistic Quantum Information (RQI) (see Ref.\cite{rmp} for a review) has shed new light on this issue. The insight from RQI is the novel observer-dependent character of quantum correlations like entanglement. For a bipartite entangled system in flat space, this means \cite{acc} an accelerated observer would experience decrement of quantum entanglement he shares initially with an inertial partner due to the celebrated Unruh effect \cite{unruh}. Such kind of environmental decoherence has later been generalized to curved background \cite{bh1}. For a static observer nearby the black hole, a degradation of quantum correlations provoked by Hawking radiation from event horizon would be detected. The entanglement produced in the formation of a black hole has also been studied and provides a quantum information resource between the field modes falling into the black hole and those radiated to infinity. By imposing proper final-state boundary conditions at the singularity \cite{finalstate}, this non-locality could transmit information outside the event horizon via a teleportation-like process and restore the unitarity of black hole evaporation process. On the other hand, it has been emphasized \cite{telep} that even the standard teleportation protocol \cite{teleportation} is highly non-trivial in the RQI framework. In flat space, as a result of entanglement degradation, the fidelity of a teleportation process would also suffer a reduction for an observer with uniform acceleration.  Moreover, quantum teleportation process in a black hole background was also investigated \cite{telep2} and it was shown that the fidelity is considerably reduced for the fixed observer near the horizon. An analogues experiment using sonic black hole is proposed \cite{telep3} to test this phenomenon in a suitable laboratory setting. 

In this Letter, we investigate a quantum teleportation process in de Sitter space, which idealizes the inflation epoch of early universe and plays a fundamental role in quantum gravity theory (see Ref. \cite{ds} for a review). We propose a protocol to teleport an unknown qubit $|\psi\ra$ from a conformal observer Alice to her inertial partner Bob while they initially share a Bell state. More specifically, we investigate two schemes of our protocol, (i) one using free scalar modes from which a qubit can be truncated \cite{truncation}, and (ii) another scheme utilizing moving cavities to store the field modes related with respective observers.  

In the scheme (i), unlike the standard teleportation protocol, we will show that the decoherence, provoked by the Gibbons-Hawking effect associated with Bob's cosmological horizon \cite{gw}, would reduce the fidelity of the teleportation process. While de Sitter space provide the best scenario to the so-called trans-Planck problem \cite{transplc}, we also discuss its influence on our teleportation scheme from the existence of some fundamental scales (like Planckian or even stringy). With a cutoff on physical momentum of field mode at Planck-scale, comparing with the standard Bunch-Davies choice, we calculate the extra modifications to the fidelity of the teleportation process from the possible high-energy new physics.

In the scheme (ii), since the field modes are localized in cavities by the reflecting mirrors, the entanglement between the cavities could be protected from the degradation once it has been prepared \cite{comment}. However, as we will show that, even using of localized cavity states could save the protocol from the Gibbons-Hawking radiation, the ability to entangle the different cavity mode is still decreased by the curvature of spacetime, that results a similar qualitatively dependence as in (i) between the fidelity and the Hubble parameter $H$.

%Review QFT in dS

To proceed, we first recall the thermal feature of quantum field theory in de Sitter space. Consider the mode expansion of a free scalar field in de Sitter space
\begin{equation}
\phi(x)=\sum_{k}[a_{k}\phi_{k}(x)+a^{\dag}_{-k}\phi_{-k}^{*}(x)]
\label{modeboson}
\end{equation}
The vacuum state, which respects the spacetime isometries, is defined by $a_{k}|vac\ra=0$. To specify the mode functions $\phi_{k}(x)$, the coordinate systems affiliated to different observers should be employed prior to solve the field equation. 

\begin{figure}[hbtp]
\includegraphics[width=.35\textwidth]{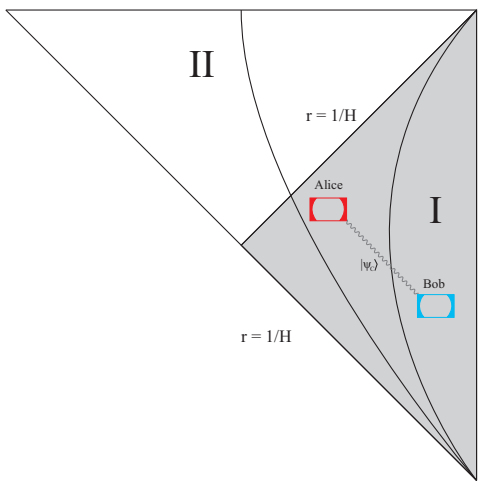}
\caption{The Carter-Penrose diagram of de Sitter space. A teleportation scheme between the conformal observer Alice and her inertial partner Bob has been illustrated. Both observers share a Bell state when they coincide initially. Since the information loss associated with Bob's cosmological horizon at $r=1/H$ (or the entanglement decrease during entangling cavities), the fidelity of teleporting a qubit $|\psi_C\ra$ would be suppressed. }
\label{telep-ds}
\end{figure}

A conformal observer in de Sitter space adopts the planar coordinates which reduce the spacetime metric to
\begin{equation}
ds^2=\frac{1}{(H\eta)^{2}}(d\eta^{2}-d\rho^{2}-\rho^{2}d\Omega^{2})
\label{confmetric}
\end{equation}
where $\eta=-e^{-Ht}/H$ is conformal time, and the coordinates cover the upper right triangle of the Carter-Penrose diagram (both regions I and II), as depicted in Fig.\ref{telep-ds}.

Since the space undergoes an accelerated expansion, it follows that the wavelength of field mode could become arbitrarily small if one goes backwards in $\eta$ long enough, where any distinction between de Sitter space and Minkowski space could be safely ignored. Therefore,  an essentially unique Bunch-Davies vacuum $a_{k}(\eta)|0,\eta\ra=0$ could be defined, by requesting it approaching the conformal vacuum of Minkowski space in the limit $\eta\rightarrow-\infty$. 

However, the existence of the fundamental scales, where the quantum gravitational effects become unignorable, prevents us from following a mode back unlimited (or equivalently, to the arbitrary short distance) \cite{transplc}. Imposing a reasonable cutoff on physical momentum as $p=ka(\eta)=\Lambda$,  the latest time with quantum gravity dominant is $\eta_{0}=-\frac{\Lambda}{Hk}$, where $\Lambda$ refers to the Planck energy scale. This results a modified vacuum state of the conformal observer as $a_{k}(\eta_{0})|0,\eta_{0}\ra=0$, which is in general different from the Bunch-Davies choice and could be formally realized as a squeezed state
\begin{equation}
|0,\eta_{0}\ra=S|0^{\infty}\ra
\label{bc2}
\end{equation}
where the superscript $\infty$ indicates the Bunch-Davies choice and throughout. Without a complete theory of quantum gravity, this new vacuum of conformal observer can provide a typical signature of Planck-scale physics. For instance, it was shown (see \cite{alpha} and the references therein) that the inflation power spectrum $P(k)\sim\la|\phi_{k}|^{2}\ra$ with respect to the new vacuum would be modified as $\Delta P(k)/P(k)=\frac{H}{\Lambda}\sin\frac{2\Lambda}{H}$, which is expected to be observed in the WMAP or Planck satellite experiments.

More ambitious view is that above argument indeed provides a one-parameter family of vacua with the $\Lambda$ predicted by various quantum gravity theories, e.g. an energy scale interpolated between Planckian and stringy scales. Equivalently, this leads the so-called $\alpha$-vacua which have been known for a long time \cite{vac1}.

Introducing the new mode basis related to the Bunch-Davies one by the Mottola-Allen (MA) transformation
\begin{equation}
\phi^{\alpha}_{k}(\eta,\vec{x})=N_{\alpha}[\phi_{k}^{\infty}(\eta,\vec{x})+e^{\alpha}\phi_{-k}^{\infty^{*}}(\eta,\vec{x})]
\end{equation}
where $\alpha$ is an arbitrary complex number with Re$\alpha<0$, $N_{\alpha}=1/\sqrt{1-e^{\alpha+\alpha^{*}}}$. The one-parameter family of vacua is defined as $a^{\alpha}_{k}|0^{\alpha}\ra=0$, where
\begin{equation}
a^{\alpha}_{k}=N_{\alpha}[a_{k}^{\infty}-e^{\alpha^{*}}a_{-k}^{\infty\;\dag}]
\label{MAboson}
\end{equation}
are the corresponding annihilation operators. These $\alpha$-vacua preserve all $SO(1,4)$ de Sitter isometries, and clearly include the Bunch-Davies vacuum as one element since $a_{k}^{\alpha}\rightarrow a_{k}^{\infty}$ if Re$\alpha\rightarrow-\infty$. Moreover, the condition (\ref{bc2}) can now be explicitly resolved as
\be
|0^\alpha_k\ra=\exp[\alpha(a^{\infty\;\dag}_ka^{\infty\;\dag}_{-k}-a^{\infty}_{-k}a^{\infty}_{k})]|0^\infty_k\ra
\label{bcboson}
\ee

In a realistic model, the value of $\alpha$ could be strictly constrained. First, for a theory consistent with $CPT$-invariance, $\alpha$ should be real. Therefore, we henceforth adopt $\alpha=$Re$\alpha$ for simplicity. On the other hand, rather than the Bunch-Davies choice, if a non-trivial $\alpha$-state ($\alpha\neq-\infty$) is chosen as an alternative initial state of inflation, the modified power spectrum of inflationary perturbations requires that \cite{transplc} $e^\alpha\sim\frac{H}{\Lambda}$.

It was shown \cite{gw} that the vacuum state defined by the conformal observer would be nonempty in the view of a static observer. While the Bunch-Davies vacuum appears thermal with the Gibbons-Hawking temperature $T=H/2\pi$, it is clear that these $\alpha$-vacua would exhibit non-thermal feature encoding the quantum gravitational corrections.

In terms of the static coordinates, de Sitter metric becomes
\be
ds^2=(1-r^2H^2)dt^2-(1-r^2H^2)^{-1}dr^2-r^2d\Omega^2
\label{static}
\ee
where $t$ is the cosmic time. The coordinates only cover the region I in Fig. \ref{telep-ds}, half of the planar coordinates dose. The hypersurface on $r = 1/H$ is a cosmological horizon for an observer situated at $r = 0$. Since the existence of the Killing vector $\partial_t$, a static vacuum $|0^S\ra$ could be defined unambiguously. To analyze the thermality of this vacuum, we employ the particular useful Painlev\'e coordinates \cite{painleve}, which reduce the metric (\ref{confmetric}) into
\be
ds^2=(1-H^2r^2)dt'^2+2Hrdrdt'-dr^2-r^2d\Omega^2
\ee
covering both regions I and II, and $t'=t+\frac{1}{2H}\ln(1-H^2r^2)$, admitting the same Killing vector as in (\ref{static}) throughout the whole space but change its character from timelike to spacelike upon passing through the horizon. The most attractive feature that we utilized is that the coordinates admit the same Bunch-Davies vacuum, while the out-state can be split into a tensor product $|I\ra\otimes|II\ra$, with $|0^{I}\ra\rightarrow|0^S\ra$,  after tracing over the degrees in region II unaccessible for the static observer in region I.  Denoting by $a_k^I$ and $a_{-k}^{II}$ the particle annihilation operators in regions I and II, we can relate them with those in the conformal framework by the Bogoliubov transformations
\be
a^{\infty}_k=\cosh r a_k^{I}-\sinh r a_{-k}^{II\;\dag}
\label{bogoboson}
\ee
With the squeezing operator $S(r)=\exp[r(a_k^{I\dag} a_{-k}^{II\dag}-a^{II}_ka_{-k}^{I})]$, the Bunch-Davies vacuum for the conformal observer can be realized as a squeezed state in the view of inertial observer
\be
|0^{\infty}_k\ra=\mbox{sech}\;r\sum_{n=0}^\infty\tanh^nr|n_k^{I};n_{-k}^{II}\ra
\label{2mode}
\ee
where $\tanh^2r=\exp(-2\pi |k|/H)$ obtained from the Gibbons-Hawking effect. This means that particles are created in pairs on either side of event horizon, and only the one in region I could be detected as de Sitter radiation by an inertial observer.

For the general $\alpha$-vacua, deviations from thermality should be included. From (\ref{bcboson}) and (\ref{bogoboson}), it follows that
\be
|0^{\alpha}_k\ra=\sqrt{1-\tanh^2r\Delta^2}\sum_{n=0}^\infty\tanh^nr\Delta^n|n_k^{I};n_{-k}^{II}\ra\label{bosealpha}
\ee
where
\be
\Delta\equiv\frac{1+e^\alpha\tanh^{-1} r}{1+e^\alpha\tanh r}=\frac{1+e^{\alpha+\pi |k|/H}}{1+e^{\alpha-\pi |k|/H}}
\label{bosoncorrec}
\ee
As $\alpha\rightarrow-\infty$, these corrections can be neglected and a pure thermal de Sitter radiation associated with the Bunch-Davies choice is recovered.
The one-particle excitation in $\alpha$-vacua can also be obtained as
\be
|1^{\alpha}_k\ra=\Big[1-\tanh^2r\Delta^2\Big]\sum_{n=0}^\infty \tanh^nr\Delta^n\sqrt{n+1}|n_k^{I}+1;n_{-k}^{II}\ra
\label{onealpha}
\ee

%Fidelity of teleportation, free mode

We now investigate our teleportation protocol in de Sitter space. As stated in the introduction, the scheme (i) of the protocol utilizes the free field modes with infinite levels, from which a a qubit can be truncated \cite{truncation}. We work with so-called dual rail encoding which is used in linear-optical quantum computing schemes and the experimental scheme of quantum teleportation protocol \cite{linopt}.

The bipartite system contains two conformal observers Alice and Bob, sharing a Bell state when they coincide. Suppose that each one supports two orthogonal modes, with the same frequency, labelled $A_i$ , $B_i$ with $i = 1, 2$, the total state of Alice and Bob encodes a two-qubit Bell state
\be
|\beta_{AB}\ra=\frac{1}{\sqrt2}(|\textbf{0}_A\ra|\textbf{0}_B\ra+|\textbf{1}_A\ra|\textbf{1}_B\ra)
\label{maxentg}
\ee
where the dual-rail basis is $|\textbf{0}_A\ra=|1^\alpha_{A_1}\ra|0^\alpha_{A_2}\ra$, $|\textbf{1}_A\ra=|0^\alpha_{A_1}\ra|1^\alpha_{A_2}\ra$, with similar expression for Bob. Here the superscript $\alpha$ indicates a general choice of vacuum state with cutoff at certain fundamental scales in de Sitter space.

After their coincidence, we suppose Bob becomes inertial while Alice maintains her comoving motion with respect to conformal time $\eta$. As we analyzed before, for the inertial observer Bob, the vacuum state chosen by Alice becomes thermal and can be written as the two-mode squeezed states (\ref{bosealpha}) and (\ref{onealpha}). To establish a teleportation protocol between Alice and Bob, which means to teleport an unknown state from conformal detector Alice to now inertial detector Bob, Alice possesses an additional qubit $|\psi_C\ra=\alpha|\textbf{0}_C\ra+\beta|\textbf{1}_C\ra$ in dual-rail basis. The input state to system is then $|\Psi\ra=|\psi_C\ra|\beta_{AB}\ra$, which can be expanded in the Bell basis associated with cavities $A$ and $C$.  If Alice makes a joint projective measurement on her two logical qubits with the result $|i\ra\otimes|j\ra, i, j\in\{0, 1\}$, the full state could be projected into
\be
|\Psi\ra=(|i\ra\otimes|j\ra)_{AC}\otimes|\phi_{ij}\ra_B
\ee
where Bob's state is
\be
|\phi_{ij}\ra_B=x_{ij}|\textbf{0}_B\ra+y_{ij}|\textbf{1}_B\ra
\label{cstate}
\ee
with the coefficients given by $(x_{00},y_{00})=(\alpha,\beta)$, $(x_{01},y_{01})=(\beta,\alpha)$, $(x_{10},y_{10})=(\alpha,-\beta)$, $(x_{11},y_{11})=(-\beta,\alpha)$. Depending on the results of measurement $\{i,j\}$ sent from Alice by classical channel, Bob can recover the unknown state by applying a proper unitary transformation on his qubit and complete the protocol in his local frame. The main point is, such a teleportation process in de Sitter space should be indeed influenced by the local Gibbons-Hawking radiation detected by Bob. Moreover, additional modification from the existence of minimal fundamental scale should also be taken into account if one starts from general  conformal vacuum like (\ref{bosealpha}) rather than the Bunch-Davies choice.

The way to probe the deviations of our scheme from the standard teleportation process in flat space is to calculate the fidelity of the teleported state of Bob. We propose that Alice should not yet cross Bob's event horizon at $r=1/H$ when she sends her measurement result out, otherwise Bob would never receive it via a classical channel. Since Bob has no access to field modes beyond his cosmological horizon, his state would be projected into a mixed one by tracing over the states in region II. From (\ref{bosealpha}), (\ref{onealpha})  and (\ref{cstate}) we have 
\bea
&&\rho_{ij}^I=\sum_{k,l=0}^\infty\ _{II}\la k,l|\phi_{ij}\ra_B\la\phi_{ij}|k,l\ra_{II}\no\\
&&=(1-\tanh^2r\Delta^2)^3\sum_{n=0}^\infty\sum_{m=0}^n\big[(\tanh^2r\Delta^2)^{n-1}[(n-m)|x_{ij}|^2\no\\
&&+m|y_{ij}|^2]|m,n-m\ra_I\la m,n-m|+(x_{ij}y^*_{ij}\tanh^{2n}r\Delta^{2n}\no\\
&&\times\sqrt{(m+1)(n-m+1)}\no\\
&&\times |m,n-m+1\ra_I\la m+1,n-m|+h.c.)\big]
\label{bobstate}
\eea
which can also be written in block diagonal form as $\rho_{ij}^I=\sum_{n=0}^\infty p_n\rho_{ij,n}^I$, with the coefficients 
\bea
p_0&=&0\ \ \ ,\ \ \ p_1=(1-\tanh^2r\Delta^2)^3 ,\no\\
p_n&=&(\tanh^2r\Delta^2)^{n-1}(1-\tanh^2r\Delta^2)^3
\eea

\begin{figure}[hbtp]
\includegraphics[width=.45\textwidth]{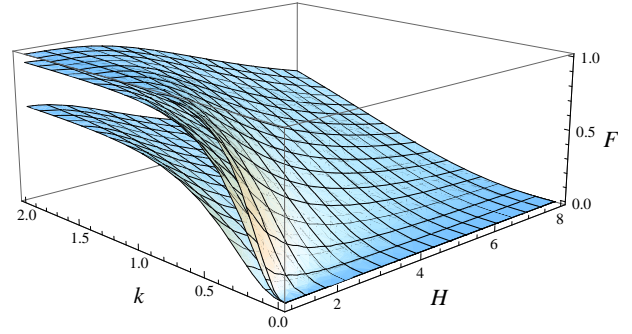}
\caption{The fidelity of teleportation in de Sitter space with $\alpha=-5,-4,-1$.}
\label{fidelity-ds}
\end{figure}

It should be emphasized that, by applying a proper unitary operation, Bob can only turn his state (\ref{bobstate}) into a region I analogue of the unknown state from Alice, like  $|\psi_I\ra=\alpha|\textbf{0}_I\ra+\beta|\textbf{1}_I\ra$, expanded in the restricted rail basis $\{|\textbf{0}_I\ra,|\textbf{1}_I\ra\}$. This deviation from the standard teleportation protocol can be measured by the fidelity defined as
\be
F^I\equiv\mbox{Tr}_I(|\psi_I\ra\la\psi_I|\rho^I)=\la\psi_I|\rho^I|\psi_I\ra=(1-\tanh^2r\Delta^2)^3
\label{fidelity}
\ee
which indicates the probability that the Bob's state $|\psi_I\ra$ will pass a test to identify it as the desired teleported state $|\psi_C\ra$. Recall the relation $\tanh^2r=\exp(-2\pi |k|/H)$ and (\ref{bosoncorrec}), above result can be depicted as in Fig.\ref{fidelity-ds}.

Our first observation is that, the fidelity of the teleportation process in de Sitter space is closely related to the Hubble parameter $H$ (or the curvature radius $l^2\equiv1/H^2$ of spacetime). As we demonstrated before, this phenomenon roots from the information loss via de Sitter radiation detected by inertial observer Bob, similar to the Unruh effect for accelerated frame in flat space or Hawking radiation from black hole. For the Bunch-Davies state which has $\alpha=-\infty$ with pure thermal spectrum, we could reach an conclusion compatible to the remarkable results of \cite{telep,telep2} where the Hubble parameter $H$ is replaced by the acceleration of the Unruh detector or the surface gravity of black hole.

Moreover, we also observe that, if one consider the possible cutoff at fundamental scales, the resulting quantum gravitational effect could be encoded in the pattern of fidelity evolution for our teleportation scheme. Comparing with the standard Bunch-Davies choice, the fidelity is suppressed for all $\alpha$-vacua choice with $\alpha\neq-\infty$. In a realistic model \cite{alpha} with $H\sim10^{14}$GeV and $\Lambda\sim10^{16}$GeV, the typical value of $\alpha$ can be estimated to be $\alpha\sim-4$. By proper tuning of the remaining parameters, this choice could result in a significantly modification in the degradation pattern of fidelity for Bob.

%localized mode

We now turn to the scheme (ii) which utilizes cavity to store the qubit for each observer. The reflecting mirrors impose the boundary conditions on the field, which result a localized mode restricted inside the cavity. While the exact solutions of field equation in curved background are highly involved, for simplicity, we demonstrate our model in a 2-dimensional de Sitter space.

We assume Alice possesses cavity $A$ described in (\ref{confmetric}) with two mirrors position at $z_i$ and the length $L=| z_2-z_1|$. The solution for the scalar field equation $(\Box+m^2)\phi=0$ which satisfied the Dirichlet boundaries $\phi(z_2)=\phi(L-z_2)=0$ is
\be
\phi^\infty_{A,k}(\eta,z)=N_A\sqrt{\pi\eta}H^{(2)}_\nu(k_n\eta)\sin[k_n(z-z_1)]
\ee
realizing the Bunch-Davies vacuum, where $k_n=\frac{n\pi}{L}$ and $H^{(2)}_\nu$ is the Hankel function with the order $\nu=(\frac{1}{4}-m^2H^{-2})^{1/2}$. The further simplification could be made for ignoring the tiny mass, then a general mode related with $\alpha$-vacua is% \cite{dic}
\be
\phi^\alpha_{A,k}(\eta,z)=\frac{N_A}{\sqrt{k_n(1-e^{2\alpha})}}\sin[k_n(z-z_1)]\big(e^{-ik_n\eta}+e^\alpha e^{ik_n\eta}\big)
\label{bdsolution}
\ee 

Similarly, the localized mode in Bob's cavity $B$ situated at $r_1=0$ could be solved in static coordinates by the hypergeometric function \cite{solution}. Within the same limitation, we have
\be
\phi_{B,k}(t,r)=\frac{N_B}{\sqrt{k'_n}}\sin\Big[\frac{k'_n}{H}\ln\sqrt{\frac{(1+Hr)(1-HL')}{(1-Hr)(1+HL')}}\bigg]e^{-ik'_nt}
\label{staticsolution}
\ee
where $k'_n=2n\pi H/\ln\frac{1+HL'}{1-HL'}$ and the static vacuum has been admitted. By the transformations between different charts (\ref{confmetric}) and (\ref{static})
\be
\eta=-\frac{e^{-tH}/H}{\sqrt{1-r^2H^2}}\qquad,\qquad z=r\frac{e^{-tH}}{\sqrt{1-r^2H^2}}
\label{atomcoordtransf}
\ee
the cavities' length in different observers' view could be related by $L'=L/\sqrt{1+L^2H^2}$. 

Since we assume perfectly reflecting mirrors, the field modes inside a cavity are protected and the detector should not be kicked \cite{cavity}. As the resource for quantum information tasks, the entanglement between the cavities now should be maintained once it was prepared. By this meaning, the teleportation protocol could be saved from Hawking radiation. However, since entangling cavities is a very non-trivial process in curved background, the ability to prepare the entangled state between Alice and Bob would result a curvature-dependence in protocol.    

We generalize the robust scheme in flat space \cite{entangcavity,movingcavity} to prepare the entangling cavities in the curved de Sitter background, which employees a two-level excited atom, passing through cavities $A$ and $B$ in their ground states. After subsequently measurement on the atom, a photon has been emitted into one of the cavities if the atom is found in its ground state. The interaction between the atom and cavity modes is described by the Hamiltonian
\be
\hat{H}_I=\hat{m}(\tau)[\epsilon_A(\tau)\phi_A(x(\tau))+\epsilon_B(\tau)\phi_B(x(\tau))]
\ee
where $\tau$ is atom's proper time and the monopole operator is $\hat{m}(\tau)=\sigma^+e^{-i\Omega\tau}+h.c$. The switching functions $\epsilon_A(\tau)$ and $\epsilon_B(\tau)$ model the effective interaction time for the atom passes through the length of cavity. By mode functions (\ref{bdsolution}), (\ref{staticsolution}), the final entangled state between Alice and Bob's cavities can be given 
\bea
|\tilde{\beta}_{AB}\ra&=&-i\int(\epsilon_A\phi^\alpha_Ae^{-i\Omega\tau}+\epsilon_B\phi_Be^{-i\Omega\tau})|0^\alpha_A\ra|0^S_B\ra d\tau\no\\
&=&\sum_k(C_A\hat{a}^{\alpha\;\dag}_k+C_B\hat{a}^{I\;\dag}_k)|0_A^\alpha\ra|0^S_B\ra\no\\
&=&\sum_k(C_A|1_A^\alpha\ra|0^S_B\ra+C_B|0_A^\alpha\ra|1^S_B\ra)
\label{entangling}
\eea
which depends on the spacetime curvature in terms of $H$, and clearly is not a maximally entangled state as (\ref{maxentg}). Once the entangled state is prepared, same local measurements could be made as in scheme (i). However, while the localized cavity states have been protected from the Gibbons-Hawking radiation, the degradation of entanglement during entangling cavity results an imperfect teleported state for Bob, similar as (\ref{cstate}) but with the coefficients $(x_{00},y_{00})=C_B(\beta,\alpha)$, $(x_{01},y_{01})=C_A(\alpha,\beta)$, $(x_{10},y_{10})=C_B(-\beta,\alpha)$, $(x_{11},y_{11})=C_A(\alpha,-\beta)$. Therefore, the fidelity loss in this scheme depends on our ability in preparing the entangled state. 

We note that the Planckian modification has been included in the possibility amplitude $C_A$. To calculate it, we specify the atom's trajectory,  $x(\tau)=(\eta(\tau),z(\tau))=(\tau,Z)$ in conformal coordinates, and can be expressed in the static coordinates as
$
(t(\tau),r(\tau))=\big(-\frac{1}{H}\ln[H^2(\tau^2-Z^2)], -\frac{Z}{H\tau}\big)
$, 
where $Z=z_1+\frac{L}{2}$ is the spatial location of the atom. For Alice's cavity, we choose a Gaussian-like switching function $\epsilon_A=\epsilon e^{-(\tau-\eta_A)^2/w^2}$, where $\eta_A$ denotes the time that atom passing through the cavity center, and $w$ depends on both the cavity geometry and the atom's transverse velocity. From (\ref{bdsolution}), (\ref{entangling}), we have
\bea
C_A&=&-i\epsilon w N_A\sqrt{\frac{\pi}{k_n(1-e^{2\alpha})}}\sin\big(\frac{n\pi}{2}\big)\no\\
&&\hspace{-40pt}\bigg[\exp\bigg(-\frac{\delta_-^2w^2}{4}+i\delta_-\eta_A\bigg)+\exp\bigg(-\frac{\delta_+^2w^2}{4}-i\delta_+\eta_A+\alpha\bigg)\bigg]\no\\
\eea
where $\delta_\pm=k_n\pm\Omega$. As $\alpha\rightarrow-\infty$, the Bunch-Davies choice is recovered, a counterpart of the remarkable results in flat space \cite{movingcavity}.

%Summary and discussion

In summary, we have demonstrated that the fidelity loss of a teleportation protocol in de Sitter space is unavoidable, either due to the Gibbons-Hawking radiation in scheme (i), or the imperfection in entangling different cavities in scheme (ii). With a cutoff at Planck scale, the possible Planckian physics cause some extra modification on the fidelity in both schemes of our teleportation protocol.

Since the important role of Planck-scale physics in early universe, we hope that those discussions on the RQI process in cosmological background \cite{cosentang} may shed some light on our understanding of quantum fluctuation decoherence during/after inflation. Even it takes a certain amount of foolhardiness to do that directly, we can at least simulate these Planckian modifications by analogue gravity experiments, like using ion trap. For instance, in detector picture \cite{ion}, to simulate Planckian modification, the detector's response function represents the probability for an ion excited by interaction with the field. The ion analogue of the Wightman function $\la\phi_m(\xi)\phi_m(\xi')\ra$ should be evaluated in some motional-state\cite{ionsqueeze}, since the $\alpha$-vacua can be interpreted as squeezed states over the Bunch-Davies vacuum state.

\section*{Acknowledgements}

J. F. has greatly benefited from conversations with Cheng-Yi Sun. This work was supported by NSFC, 973 program (2010CB922904). W.L.Y. acknowledges the support of NSFC (Grant Nos. 11075126 and 11031005). Y.Z.Z. acknowledges the support of the Australian Research Council through DP110103434.

\end{document}